\documentclass[11pt,a4paper,english]{article}
\usepackage{epsfig}
\usepackage{float}
\usepackage{amsmath}
\usepackage{graphicx}

\newcommand{\ba}{\begin{array}}
\newcommand{\ea}{\end{array}}
\newcommand{\bd}{\begin{displaymath}}
\newcommand{\ed}{\end{displaymath}}
\newcommand{\bi}{\begin{itemize}}
\newcommand{\ei}{\end{itemize}}
\newcommand{\benu}{\begin{enumerate}}
\newcommand{\eenu}{\end{enumerate}}
\newcommand{\be}{\begin{equation}}
\newcommand{\ee}{\end{equation}}
\newcommand{\bea}{\begin{eqnarray}}
\newcommand{\eea}{\end{eqnarray}}

\newcommand{\bad}{\begin{array}{ccc}}

\newcommand{\beq}{\begin{equation}}
\newcommand{\eeq}{\end{equation}}

\usepackage{textcomp}

\makeatother

\usepackage{babel}

\begin{document}

\begin{center}

{\bf{\large Addendum: Neutrino Mass Hierarchy Determination Using Reactor 
Antineutrinos}}



\vspace{0.4cm}

Pomita Ghoshal$\mbox{}^{a)}$
\footnote{pomita.ghoshal@gmail.com},
S.T. Petcov$\mbox{}^{b,c)}$
\footnote{Also at: Institute of Nuclear Research and Nuclear Energy, 
Bulgarian Academy of Sciences, 1784 Sofia, Bulgaria.}

\vspace{0.2cm}

$\mbox{}^{a)}${\em Physical Research Laboratory, Navrangpura, 
Ahmedabad, India.\\
}

\vspace{0.1cm}
$\mbox{}^{b)}${\em SISSA and INFN-Sezione di Trieste, Trieste, Italy\\}

\vspace{0.1cm}
$\mbox{}^{c)}${\em Kavli IPMU, University of Tokyo, Tokyo, Japan.\\}

\end{center}

\begin{abstract}
We update our study of 
neutrino mass hierarchy determination
using a high statistics reactor $\bar{\nu}_e$ experiment 
in the light of the recent evidences of a relatively 
large non-zero value of $\theta_{13}$ from the Daya Bay and
RENO experiments. We find that there are noticeable modifications 
in the results, which allow a relaxation in 
the detector's characteristics, such as 
the energy resolution and exposure, 
required to obtain a significant 
sensitivity to, or to determine, 
the neutrino mass hierarchy 
in such a reactor experiment.
\end{abstract}

\section{Introduction}

  Determining the type of neutrino mass spectrum, which can be with normal 
or inverted ordering (NO or IO) or hierarchy  (see, e.g., \cite{PDG10}), 
is one of the most pressing and challenging problems of future research in 
neutrino physics. The recently measured relatively large value of 
the angle $\theta_{13}$ of the Pontecorvo, Maki, Nakagawa, Sakata 
(PMNS) neutrino mixing matrix in the Daya Bay 
\cite{An:2012eh} and RENO \cite{RENOth13} experiments
\footnote{
{\bf The angle $\theta_{13}$ was found to be different 
 from zero, respectively at $5.2\sigma$ and $4.9\sigma$ 
in the Daya Bay and RENO experiments. Subsequently, 
the Double Chooz \cite{DChoozNu12} and T2K \cite{T2KNu12} 
experiments reported 
$3.1\sigma$ and $3.2\sigma$ evidences for a nonzero  
value of $\theta_{13}$.}
}
opens up the possibility of the neutrino mass hierarchy 
determination in an experiment with reactor $\bar{\nu}_e$. 
This possibility was discussed first 
in \cite{PPiai02} and later was further investigated in 
\cite{SCSPMP03,Hano12,YWang08,Ghoshal:2010wt} 
(see also \cite{Schonert:2002ep}).
It is based on the observation that for 
$\cos2\theta_{12} \neq 0$ and $\sin\theta_{13}\neq 0$, 
$\theta_{12}$ being the solar neutrino mixing angle 
(see, e.g., \cite{PDG10}),
the probabilities of $\bar{\nu}_e$ survival 
in the cases of NO (NH) and IO (IH) spectra 
differ \cite{PPiai02,BiNiPe02}: 
$P^{NH}({\bar \nu_e}\to{\bar \nu_e})\neq
P^{IH}({\bar \nu_e}\to{\bar \nu_e})$, and 
$|P^{NH}({\bar \nu_e}\to{\bar \nu_e}) - 
P^{IH}({\bar \nu_e}\to{\bar \nu_e})| \propto 
\sin^22\theta_{13} \cos2\theta_{12}$.
For sufficiently large $|\cos2\theta_{12}|$ and 
$\sin^2\theta_{13}$ and a baseline of 
several tens of kilometers, this difference 
in the  $\bar{\nu}_e$ oscillations
leads, in principle, to an observable
difference in the deformations 
of the spectrum of $e^+$ \cite{PPiai02}, 
produced in the inverse beta-decay reaction 
$\bar{\nu}_e + p \rightarrow e^+ + n$ 
by which the reactor $\bar{\nu}_e$ are detected. 

In the present Addendum we re-evaluate 
the potential of the reactor $\bar{\nu}_e$ 
experiments for determination of the neutrino mass 
hierarchy using the Daya Bay and RENO data on $\theta_{13}$.
Such a re-evaluation is necessary since 
$\sin^2\theta_{13}$ was measured with a relatively high 
precision in the Daya Bay and RENO experiments 
and found to have a relatively large value.
We expect the latter to lead to less 
demanding, than previously estimated,
characteristics of the $\bar{\nu}_e$ detector, 
required for getting information about the type of 
the neutrino mass spectrum.

  We perform the analysis uing the methods descibed in detail 
in \cite{Ghoshal:2010wt}. 
We assume the experiment is 
performed with a KamLAND-like 
(see, e.g., \cite{Inoue:2004wv}) 
10 kT detector (planned, e.g., within the project 
Hanohano \cite{Dye:2006gx}), 
located at $L = 60$ km from a reactor $\bar{\nu}_e$ source, 
having a power of $\sim 5$ GW. 
As in \cite{Ghoshal:2010wt} (see also \cite{SCSPMP03}),
the threshold of the visible energy used is set to 
$E_{visth} = 1.0$ MeV.
 As is well known,  for the experimentally determined 
values of the solar and atmospheric neutrino mass 
squared differences, which we give below, 
the optimal baseline of the experiment of interest 
is approximately 60 km (see, e.g., \cite{PPiai02,Hano12}).
We present results also for the shorter non-optimal 
baseline of $L = 30$ km.
For the reactor angle $\theta_{13}$,
we use the results of the Daya Bay 
experiment \cite{An:2012eh}:
\begin{equation}
 \sin^22\theta_{13} = 0.092 \pm 0.016 \pm 0.005\,,~~
 0.04 \leq \sin^22\theta_{13} \leq 0.14\,,~3\sigma\,.
\label{DBayth13}
\end{equation}
%

 In what concerns the other oscillation parameters 
which enter into the expressions for the 
reactor $\bar{\nu}_e$ survival probabilities in the 
cases of NO and IO spectra, the solar and atmospheric 
neutrino mass squared differences, 
$\Delta m^{2}_{\odot}\equiv \Delta m^{2}_{21}$ 
and $\Delta m_A^2 \equiv \Delta m^{2}_{31} \cong 
\Delta m^{2}_{32}$, and the solar neutrino mixing 
angle, $\theta_{12}$, we use the values 
obtained in the global analysis of the neutrino oscillation data, 
including the data from the Daya Bay and RENO experiments, 
performed in \cite{Fogli:2012ua}. It follows from the results 
obtained in  \cite{Fogli:2012ua}, in particular, 
{ that  we have $\cos2\theta_{12} \geq 0.28$ at $3\sigma$ }.

Since the sensitivity to the neutrino mass hierarchy 
of a reactor ${\bar{\nu}}_e$ experiment depends 
{critically} on the value of the angle $\theta_{13}$, 
we have redone our earlier analysis \cite{Ghoshal:2010wt} 
taking into account the new data on $\sin\theta_{13}$, 
eq. (\ref{DBayth13}),
including the allowed $3\sigma$ interval of values. 
In the following Section we present our 
updated analysis and results.

{\section{Updated $\chi^2$-Analysis of the sensitivity to the 
type of the neutrino mass spectrum}}

 We perform a full $\chi^2$ analysis of the hierarchy sensitivity of 
a medium-baseline reactor ${\bar{\nu}}_e$ experiment 
with a detector of the prototype of KamLAND, choosing 
the optimal baseline of 60 km unless otherwise stated. 
The hierarchy sensitivity 
is computed by simulating an ''experimental'' event spectrum 
for a fixed ''true'' hierarchy (we choose a normal hierarchy here, 
the difference being minimal if it is chosen to be the inverted one). 
A ''theoretical'' event spectrum is simulated with the other 
or ''wrong'' hierarchy. A standard Gaussian $\chi^2$ is then 
obtained, which determines the confidence level at which 
the ''wrong'' hierarchy can be excluded.

 Our rigorous analysis involves optimizing the event 
binning to give the best sensitivity while being 
compatible with constraints of detector resolution, 
marginalizing over the neutrino parameters  
$|\Delta m^2_{\rm atm}|$ and $\theta_{13}$, 
and taking into account systematic and geo-neutrino 
uncertainties by the method of pulls 
(for further technical details of the  analysis 
see  \cite{Ghoshal:2010wt}).
We have checked in \cite{Ghoshal:2010wt} that doing a 
marginalization over  $\sin^2 \theta_{12}$ and $\Delta m^2_{21}$
over their present $3\sigma$ ranges
of $\sin^2 \theta_{12} = 0.26 - 0.36$ and 
$\Delta m^2_{21} = 
7.0\times 10^{-5}~{\rm eV^2}- 8.2\times 10^{-5}~{\rm eV^2}$ 
\cite{Fogli:2012ua} 
does not significantly affect the results on hierarchy sensitivity,
since they are relatively small variations. 
Hence, we have presented  in \cite{Ghoshal:2010wt} the final
results with the values of $\sin^2 \theta_{12}$ and 
$\Delta m^2_{21}$ fixed at their best-fits of
$\sin^2\theta_{12} = 0.31$ and 
$\Delta m^2_{21} = 7.6\times 10^{-5}~{\rm eV^2}$.
We follow the same procedure here
\footnote{
We have made use also of the results found in 
\cite{Ghoshal:2010wt} (see also \cite{SCSPMP03,Hano12})
that the inclusion of systematic and geo-neutrino 
uncertainties as well as of $\sim 1\%$
energy scale shrink/shift 
uncertainty (even if energy-dependent), 
has only a minimal effect on 
the neutrino mass  hierarchy determination.
}.

We present results for different values of $\theta_{13}$, 
the detector exposure and the energy resolution. 
As was done in \cite{Ghoshal:2010wt},  
a prior term is added to take into 
account information from other experiments on parameter 
uncertainties. We find that the uncertainties in the 
values of $|\Delta m^2_{\rm atm}|$ and $\theta_{13}$ 
play a crucial role in the sensitivity to the neutrino 
mass hierarchy, and hence the reduction in the allowed 
range of $\theta_{13}$ as well as its increased 
value aid in hierarchy determination. 
We study the effect of the detector energy resolution, 
exposure, parameter marginalization and data binning
using the new data on $\theta_{13}$, eq. (\ref{DBayth13}).

\vskip.2cm

We consider the following error ranges for the 
two marginalized parameters:
i) $|\Delta m^2_{31}|$ is allowed to vary
  in the range $2.3 \times 10^{-3} - 2.6
  \times 10^{-3}$ eV$^2$, and
ii) $\sin^2 2 \theta_{13}$ is varied from 0.04 to 0.14, to be 
consistent with the 3$\sigma$ range found in the Daya Bay 
experiment.

\vskip.2cm

\begin{figure}[t]
\centerline
{
\epsfxsize=8.5cm\epsfysize=7.0cm
\epsfbox{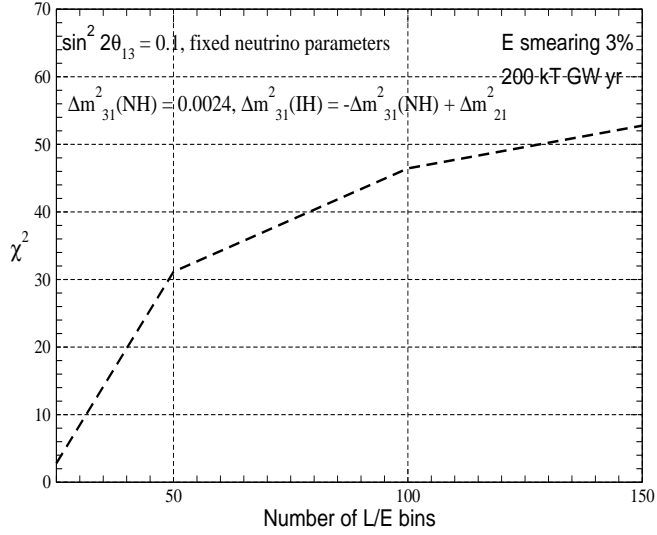}
}
\caption[]
{\footnotesize{
The hierarchy sensitivity $(\chi^2)_{stat}$ 
as a function of the number of L/E bins, 
for fixed neutrino oscillation parameters,
${\rm{\sin^2 2\theta_{13}}} = 0.1$ and detector's 
energy resolution of 3$\%$, statistics of 200 kT GW yr,
baseline of 60 km and different L/E binnings in 
the range ${\rm{L/E}} = 5-32$ km/MeV.
}}
\label{chisqvsbins}
\end{figure}

 Figure~\ref{chisqvsbins} shows the behaviour of the 
$\chi^2$ sensitivity with an increase in the bin number 
for fixed neutrino parameters and an exposure of 200 kT GW yr, 
using $\sin^2 2\theta_{13} = 0.1$, $\Delta m^2_{31}(NH) = 
 2.4\times 10^{-3}~{\rm eV^2}$, 
$\Delta_{31}(IH) = -\Delta_{31}(NH) + \Delta m^2_{21}$ 
and a detector resolution of 3$\%$, 
for different numbers of L/E bins in the range  
${\rm{L/E}} = 5-32$ km/MeV. The sensitivity is seen 
to improve dramatically with an improvement in 
the fineness of division, and the binning is 
optimized at 150 L/E bins to derive the best possible 
sensitivity while being consistent with the detector resolution. 
For 150 (100) L/E bins, the bin width in energy in the 
case we are considering is about 68 (100) keV. 

\vskip.2cm

%

  Table 1 lists the values of the hierarchy sensitivity 
$(\chi^2)_{stat}^{min}$ for different values of $\theta_{13}$ and 
the detector energy resolution, after a marginalization 
over the parameter ranges indicated above, for an exposure of
200 kT GW yr and a 150-bin analysis. 
The true $\theta_{13}$ values are chosen within the 
$3\sigma$ range allowed by the Daya Bay data. 
Prior experimental information regarding the 
other neutrino parameters is included 
in the analysis in the form of ''priors'',
using the present 1$\sigma$ error ranges 
of the respective parameters:
$\sigma(|\Delta m^2_{\rm{atm}}|)=
5$\%$ \times |\Delta m^2_{\rm{atm}}|^{\mathrm{true}}$ 
and $\sigma(\sin^2 2\theta_{13})=0.02$. 
Table 2 gives the values of the hierarchy 
sensitivity $[(\chi^2)_{stat}^{min}]_{prior}$ 
for different values of $\theta_{13}$ and the 
detector energy resolution with a parameter marginalization 
including priors, for the same values of 
detector exposure and event binning. 
The slight improvement in the results 
with the inclusion of priors is enhanced if  
a lower prospective  1$\sigma$ error of  
$\sigma(\sin^2 2\theta_{13}) = 0.01$ is considered. 
As recent reports from Daya Bay and RENO have shown, 
such an improvement
in the precision of $\theta_{13}$ is not far out 
of reach of present experiments. Moreover, 
a combined analysis of the global data 
on the angle $\theta_{13}$ performed in \cite{Combinedth13} 
already yields $\sigma(\sin^2 2\theta_{13}) = 0.013$.
%
 
\vskip.2cm

\begin{table}[t]
\begin{center} 
\begin{tabular}{| c || c | c | c | }
\hline 
 {\sf{$(\chi^2)_{stat}^{min}$}}  & {\sf {Energy resolution}} & & \\
        \hline
        \hline
{\sf {$\sin^2 2\theta_{13}^{\rm{true}}$}} & 2$\%$ & 3$\%$ & 4$\%$ \\
        \hline 
        \hline
         0.07 & 6.21  & 4.99  & 3.81   \\
         \hline
          0.1  & 12.91 & 10.41 & 7.90 \\
           \hline
           0.12 & 18.80 & 15.10 & 11.48 \\
           \hline
            \end{tabular}
            \caption[]{\footnotesize{
Values of $(\chi^2)_{stat}^{min}$
marginalized over the parameters $\theta_{13}$ and 
$|\Delta m^2_{31}|$, for 
$|\Delta m^2_{31}|^{\mathrm{true}} = 2.4\times 10^{-3}~{\rm eV^2}$,
$\sigma(|\Delta m^2_{31}|)=
5$\%$ \times |\Delta m^2_{31}|^{\mathrm{true}}$,
$\sigma(\sin^2 2\theta_{13})=0.02$,
three values of $\sin^22\theta_{13}^{\rm{true}}$ 
and three values of the detector energy resolution. The
detector exposure used is 200 kT GW yr. The baseline is set 
to 60 km. The values  of $(\chi^2)_{stat}^{min}$ 
are obtained in an analysis using 150 L/E bins 
in the range 5 - 32 km/MeV.}} 
            \label{table1}
            \end{center}
            \end{table}

In Table 3, we list 
the values of the hierarchy 
sensitivity $[(\chi^2)_{stat}^{min}]$ for 
$\sin^2 2\theta_{13}^{\mathrm{true}}=0.07$ and $0.1$, 
for 3 different values of the detector resolution 
and a scaling in the detector exposure. These 
results show the strong dependence of the 
sensitivity on the detector exposure. 
For example, a hierarchy 
sensitivity of nearly $3\sigma$ may be possible 
even for $\sin^2 2\theta_{13}^{\mathrm{true}}=0.07$
and an energy resolution of 4$\%$, with an exposure of 400 kT GW yr, 
and this would improve further with a higher detector mass/power.

 To highlight the improved sensitivities possible even for 
smaller detector exposures when $\theta_{13}$ is close to 
the present best-fit value, 
we present in Table 4 the hierarchy sensitivity  $[(\chi^2)_{stat}^{min}]$ for 
$\sin^2 2\theta_{13}^{\mathrm{true}}=0.1$ and $0.12$ for 
lower detector exposures 100 kT GW yr and 150 kT GW yr 
with 3 different values of the detector's energy
resolution. We note that even with an energy resolution of 4$\%$, 
a 2$\sigma$ sensitivity is achievable with a relatively low 
exposure of 100 kT GW yr for the indicated 
values of $\theta_{13}$. With a better energy resolution 
of 2$\%$, the sensitivity can go up to 3$\sigma$ or 
even to a higher value.

 In Table 5 we list the values of hierarchy sensitivity obtained 
for two detector exposures and three values of detector resolution 
when the baseline is chosen to be 30 km instead of 60 km. 
This table shows that the sensitivities decrease for 
the indicated shorter baseline, i.e., when the baseline 
deviates significantly from the optimal one 
of 50 - 60 km. For example, with a baseline of 30 km, 
a resolution of 2$\%$ and an exposure of 200 kT GW yr 
would be required for a hierarchy sensitivity of 
3$\sigma$ if $\sin^2 2\theta_{13}^{\mathrm{true}}=0.1$, 
while with a baseline of 60 km similar sensitivity 
is achievable with an exposure of 150 kT GW yr with 
the same detector resolution
\footnote{
{\bf{The optimal baseline for hierarchy 
sensitivity lies in the region of maximization of 
the effect of the phase 
$\Delta m_{21}^2 L/2E$ in the expression for the 
$\bar{\nu}_e$ survival probability. 
With the present error range of $\Delta m_{21}^2$,
and the peak of the reactor $\bar{\nu}_e$ event 
rate spectrum at 3.6 MeV, 
this gives an optimal baseline range 
of 55 to 64 km. Hence, the hierarchy sensitivity 
becomes worse for baselines significantly 
shorter than the indicated range}}. 
}.
\begin{table}[t]
\begin{center} 
\begin{tabular}{| c || c | c | c | }
\hline 
 {\sf{$[(\chi^2)_{stat}^{min}]_{prior}$}}  & {\sf {Energy resolution}} & & \\
        \hline
        \hline
{\sf {$\sin^2 2\theta_{13}^{\rm{true}}$}} & 2$\%$ & 3$\%$ & 4$\%$ \\
        \hline 
        \hline
       0.07 & 6.37 & 5.15 & 3.90   \\
         \hline
          0.1  & 13.17 & 10.58 & 8.05   \\
           \hline
           0.12 & 19.10 & 15.26 & 11.60 \\
           \hline 
            \end{tabular}
            \caption[]{\footnotesize{
The same as in Table \ref{table1}, but for 
$\sigma(\sin^2 2\theta_{13}) = 0.01$.
}}
            \label{table2}
            \end{center}
            \end{table}
\vskip.4cm
\begin{table}[t]
\begin{center}
\begin{tabular}{| c || c | c | c || c | c | c | }
\hline
 {\sf{$(\chi^2)_{stat}^{min}$}} & $\sin^2 2\theta_{13}^{\rm{true}}=0.07$ & 
& & $\sin^2 2\theta_{13}^{\rm{true}}=0.1$ & & \\
 \hline
 \hline
{\sf{ Detector exposure, kT GW yr}} & {\sf {Energy resolution}} & & & & & \\
        \hline
        \hline
 & 2$\%$ & 3$\%$ & 4$\%$ & 2$\%$ & 3$\%$ & 4$\%$ \\
        \hline
        \hline
         200 & 6.21  & 4.99  & 3.81 & 12.91 & 10.41 & 7.90 \\
         \hline
          400  & 12.40 & 9.98 & 7.60 & 25.80 & 20.80 & 15.78 \\
           \hline
            600   & 18.61  & 14.95  & 11.71 & 38.70 & 31.20 & 23.50 \\
             \hline
            \end{tabular}
            \caption[]{\footnotesize{
The same as in Table \ref{table1}, but for three values of the 
detector exposure and
$\sin^2 2\theta_{13}^{\rm{true}}=0.07;~0.1$.
}}
            \label{table3}
            \end{center}
            \end{table}
\vskip.4cm
\begin{table}[t]
\begin{center}
\begin{tabular}{| c || c | c | c || c | c | c | }
\hline
 {\sf{$(\chi^2)_{stat}^{min}$}} & $\sin^2 2\theta_{13}^{\rm{true}}=0.1$ & 
& & $\sin^2 2\theta_{13}^{\rm{true}}=0.12$ & & \\
 \hline
 \hline
{\sf{ Detector exposure, kT GW yr}} & {\sf {Energy resolution}} & & & & & \\
        \hline
        \hline
 & 2$\%$ & 3$\%$ & 4$\%$ & 2$\%$ & 3$\%$ & 4$\%$ \\
        \hline
        \hline
         100 & 6.50  & 5.20  & 3.98 & 9.45 & 7.57 & 5.75 \\
         \hline
          150  & 9.70 & 7.80 & 5.95 & 14.15 & 11.35 & 8.60 \\
             \hline
            \end{tabular}
            \caption[]{\footnotesize{
Values of $(\chi^2)_{stat}^{min}$
marginalized over the parameters $\theta_{13}$ and $|\Delta m^2_{31}|$
for lower detector exposures (in kT GW yr),
$\sin^2 2\theta_{13}^{\rm{true}}=0.1$ and $0.12$, for three values of 
the detector's energy resolution
and a baseline of 60 km. The results are obtained
in an analysis using 150 L/E bins in the range 5 - 32 km/MeV.
}}
            \label{table4}
            \end{center}
            \end{table}
\vskip.4cm
\begin{table}[t]
\begin{center}
\begin{tabular}{| c || c | c | c || c | c | c | }
\hline
 {\sf{$(\chi^2)_{stat}^{min}$ (30 km)}} & $\sin^2 2\theta_{13}^{\rm{true}}=0.1$ & 
& & $\sin^2 2\theta_{13}^{\rm{true}}=0.12$ & & \\
 \hline
 \hline
{\sf{ Detector exposure, kT GW yr}} & {\sf {Energy resolution}} & & & & & \\
        \hline
        \hline
 & 2$\%$ & 3$\%$ & 4$\%$ & 2$\%$ & 3$\%$ & 4$\%$ \\
        \hline
        \hline
         150 & 6.60  & 4.90  & 3.80  & 9.65 & 7.15  & 5.54 \\
         \hline
          200  & 8.79  & 6.50 & 5.05 & 12.81 & 9.48 & 7.35 \\
             \hline
            \end{tabular}
            \caption[]{\footnotesize{
Values of $(\chi^2)_{stat}^{min}$
marginalized over the parameters $\theta_{13}$ and $|\Delta m^2_{31}|$
for two values of detector exposures (in kT GW yr), for
$\sin^2 2\theta_{13}^{\rm{true}}=0.1$ and $0.12$, three values of 
the detector's energy resolution
and a baseline of 30 km. The results are obtained
in an analysis using 150 L/E bins in the range 5 - 32 km/MeV.
}}
            \label{table5}
            \end{center}
            \end{table}

%
\section{Conclusions}
%

We find that the data on the parameter $\theta_{13}$ 
from Daya Bay experiment allow us to get information or determine 
the neutrino mass hierarchy with a greater 
efficiency, than was previously estimated,
using a reactor ${\bar{\nu}}_e$ experiment:
the stringent requirements of the detector's 
energy resolution and exposure obtained in the previous studies
can be relaxed significantly. 
Since hierarchy sensitivity depends strongly on the 
the true value of $\theta_{13}$, the energy resolution
and the exposure, a relatively large 
value of ${\rm{\sin^22\theta_{13}^{true}}}$ close to 
the Daya Bay best fit of 0.092 makes it easier to achieve
hierarchy 
determination
using lower detector exposures 
and less demanding energy resolution. 

\vskip.2cm

For example, $(\chi^2)^{min}_{stat}$ for the 
``wrong'' hierarchy improves from 3.5 (1.8$\sigma$ sensitivity) for 
$\rm{sin^2 2\theta_{13}^{true}} = 0.05$ 
(close to the Daya Bay $3\sigma$ lower limit), 
an energy resolution of 2$\%$ and a detector exposure 
of 200 kT GW yr, to 12.9 (a 3.6$\sigma$ determination)
for $\rm{sin^2 2\theta_{13}^{true}} = 0.10$ (close to the Daya Bay best fit) 
for the same values of the resolution and exposure.
With this value of  $\rm{sin^2 2\theta_{13}^{true}}$, 
even an energy resolution of 4$\%$
can give a sensitivity of nearly 3$\sigma$. 

\vskip.2cm

To summarise, for the values of $\theta_{13}$ from the interval allowed at 
$3\sigma$ by the Daya Bay data, a significant hierarchy sensitivity 
is possible even with a detector energy resolution of $\sigma\sim 4\%$ 
and an exposure of 200 kT GW yr. 
For $\rm{sin^2 2\theta_{13}^{true}} = 0.10~(0.12)$ and 
an energy resolution of  $\sigma\sim 2\%$, a $3\sigma$ sensitivity 
to the neutrino mass hierarchy can be achieved with an exposure 
of  150 (100) kT GW yr. The indicated requirements on the detector 
specifications make the discussed reactor ${\bar{\nu}}_e$ experiment 
more feasible than the previous analyses have suggested.

\vskip.25in

\textbf{Acknowledgements}

The work of S.T.P. was supported in part by 
the INFN program on ``Astroparticle Physics'', 
by the Italian MIUR program on
``Neutrinos, Dark Matter and  Dark Energy in the Era of LHC'', 
by the World Premier International
Research Center Initiative (WPI Initiative), MEXT,
Japan, and by the European Union
under FP7 ITN INVISIBLES (Marie Curie Actions,
PITN-GA-2011-289442). P.G. thanks the XI Plan Neutrino Project at HRI, Allahabad, for providing financial assistance to visit HRI, during which some parts of the work were carried out.


\begin{thebibliography}{99}
\bibitem{PDG10}
K. Nakamura {\it et al.} (Particle Data Group),
J. Phys. {\bf G 37} (2010) 075021.


\bibitem{An:2012eh}
  F.~P.~An {\it et al.}  [DAYA-BAY Collaboration],
Phys. Rev. Lett. {\bf 108} (2012) 171803.

\bibitem{RENOth13}
 J.K. Ahn {\it et al.} [RENO Collaboration],
Phys. Rev. Lett. {\bf 108} (2012) 191802.

\bibitem{DChoozNu12}
I. Masaki [for the Double Chooz Collaboration],
talk at Neutrino 2012, the XXV International Conference on
Neutrino Physics and Astrophysics June 4-10, 2012, Kyoto, Japan
(available at the web-site neu2012.kek.jp).

\bibitem{T2KNu12} T. Nakaya [for the T2K Collaboration],
talk at Neutrino 2012, the XXV International Conference on
Neutrino Physics and Astrophysics June 4-10, 2012, Kyoto, Japan
(available at the web-site neu2012.kek.jp).

\bibitem{PPiai02}
S.T. Petcov and M. Piai, 
Phys. Lett. B {\bf 533} (2002) 94.

\bibitem{SCSPMP03} S. Choubey, S.T. Petcov and M. Piai,
Phys. Rev. D {\bf 68} (2003) 113006.

\bibitem{Hano12} J. Learned {\it et al.}, 
Phys. Rev. D {\bf 78} (2008) 071302;
M. Batygov {\it et al.}, arXiv:0810.2580.

\bibitem{YWang08} L. Zhan {\it et al.}, 
Phys. Rev. D {\bf 78} (2008) 111103
and Phys. Rev. D {\bf 79} (2009) 073007.

\bibitem{Ghoshal:2010wt}
  P.~Ghoshal and S.~T.~Petcov,
  JHEP {\bf 1103} (2011) 058.


\bibitem{Schonert:2002ep}
  S.~Schonert, T.~Lasserre and L.~Oberauer,
Astropart.\ Phys.\  {\bf 18} (2003) 565.


\bibitem{BiNiPe02} S.M. Bilenky, D. Nicolo and S.T. Petcov,
Phys. Lett. B {\bf 538} (2002) 77.

\bibitem{Inoue:2004wv}
  K.~Inoue,
  New J.\ Phys.\  {\bf 6} (2004) 147.

\bibitem{Dye:2006gx}
  S.~T.~Dye, E.~Guillian, J.~G.~Learned, J.~Maricic, 
S.~Matsuno, S.~Pakvasa, G.~S.~Varner and M.~Wilcox,
anti-neutrino observatory,''
Earth Moon Planets {\bf 99} (2006) 241
[hep-ex/0609041].

\bibitem{Combinedth13} P.A.N. Machado {\it et al.},
arXiv:1111.3330, to be published in JHEP.

\bibitem{Fogli:2012ua}
  G.~L.~Fogli, E.~Lisi, A.~Marrone, D.~Montanino, A.~Palazzo and A.~M.~Rotunno,
  arXiv:1205.5254 [hep-ph].


\end{thebibliography}
\end{document}